# Nonlinear beam self-maintaining effect in graded-index multimode fiber

Baofu Zhang[1], Shanchao Ma[2], Qiurun He[2], Sihua Lu[2], Wei Guo[2], Zhongxing Jiao[2, *], and Biao Wang[1, 2, *]

[1] Research Institute of Interdisciplinary Science & School of Materials Science and Engineering, Dongguan University of Technology, Dongguan 523808, China

[2] School of Physics, Sun Yat-sen University, Guangzhou 510275, China

*Corresponding author: jiaozhx@mail.sysu.edu.cn, wangbiao@mail.sysu.edu.cn.

**Abstract**

Multimode fiber systems are desirable for industrial and scientific applications. As an interesting effect for the laser beam propagation in a multimode fiber, nonlinear Kerr beam cleanup has attracted considerable research interest due to the spatial beam compressing. However, its physical mechanisms, especially the influences of input conditions on its performances, remain unclear. Here, we report a new self-organized regime for the multimode beam propagation in a graded-index multimode fiber: when the input laser has a dominant mode in which most of the laser energy is concentrated, the beam profile can be maintaining in a well-defined structure similar to the input dominant mode in nonlinear regime, while it will evolve to an irregular pattern in linear regime. The existence and universality of this nonlinear beam self-maintaining effect have been verified by the experimental and numerical data. Our results also provide evidence that nonlinear Kerr effects can be the driving mechanism and nonlinear Kerr beam cleanup is a specific case of this effect. Further research into this spatial beam shaping effect may provide a new perspective to understand other multimode fiber nonlinearities.

## 1. Introduction

In the past few years, widespread research interest in multimode optical fibers (MMFs) has been revived. On one hand, MMFs play a key role in many important applications such as space-division multiplexing [1], high-resolution imaging [2, 3], and high-power fiber lasers [4]. On the other hand, MMFs provide a convenient platform for studying complex physical systems [5, 6]. Due to the sophisticated interplay of spatiotemporal processes during beam propagation in MMFs, there are rich and complex nonlinear phenomena for investigation, including multimode solitons [7], spatiotemporal mode-locking [8] and parametric instability [9].

Among these nonlinear phenomena in MMFs, nonlinear Kerr beam cleanup (NL-KBC) is an interesting spatial beam shaping effect which arises when intense laser pulses propagate in a graded-index multimode fiber (GIMF) [10, 11]. With the increase of input laser power, the output beam profile evolves from an irregular multimode pattern in linear regime into a well-defined bell-shaped structure in nonlinear regime. Although NL-KBC has been observed and investigated in various optical systems [10-19], its physical mechanisms remain unclear.

One of the theories is based on the nonreciprocal behavior of nonlinear mode coupling [10]: due to the nonlinear Kerr effects and four wave mixing in the GIMF, energy of high-order modes (HOMs) can be irreversibly coupled into the fundamental mode during the propagation. However, some reported phenomena along with NL-KBC are difficult to be explained by the theory mentioned above. First, the impact of the number of initially excited modes have been investigated by imposing a transverse shift to the lateral position of the input fundamental Gaussian beam with respect to the core of GIMF [6, 12]. In this case, researchers indicated higher threshold or smaller efficiency for NL-KBC with the larger number of excited modes. Nevertheless, no trace of spatially compressed beam could be found in nonlinear regime when input beam position was far from the center of GIMF, which is slightly inconsistent with the nonreciprocal coupling theory. Second, NL-KBC on the $LP_{11}$ mode and other high-order linearly polarized (LP) modes have been demonstrated experimentally with the incident external angle [13-15] or by using a deformable mirror [16]. In these cases, researchers attributed the cause to a Kerr-induced index grating with a spatial geometry leading to a strong overlap with the desired mode. However, these unexpected phenomena are quite different from the nonreciprocal coupling theory in which the intermodal energy exchanges may favor the fundamental mode of GIMF.

According to these experimental results, it can be found that the generation of NL-KBC depends seriously on the input conditions. For the above phenomena, we think the most important factor is whether there is an input dominant transverse mode but not the number or the order of excited modes. Here, we propose a conjecture based on self-organization through the cooperation of disorder and nonlinearity [20]. When the input laser has a dominant mode in which most of the laser energy is concentrated, the beam profile can be maintaining in a well-defined structure similar to the input dominant mode in nonlinear regime, while it will evolve to an irregular pattern in the linear regime. In our conjecture, NL-KBC can serve as an exceptional case of this nonlinear beam self-maintaining (NL-BSM) effect in which the fundamental mode ($LP_{01}$ mode) is the input dominant mode.

In order to directly verify our conjecture about the NL-BSM and clarify its physical mechanism, our approach is to obtain high-peak-power input laser beam with dominant HOMs instead of the fundamental mode reported in most of NL-KBC experiments. In this paper, experiments with multimode laser beam propagation in GIMF are carried out to verify the existence and universality of NL-BSM. The performances and mechanisms of this nonlinear effect are also studied by numerical simulations. Here, we demonstrate a new regime of self-induced spatial beam shaping based on nonlinear energy maintaining of dominant mode and disorder-induced energy exchange of other modes.

## 2. Methods

The experiment setup is similar to the one presented in our previous work [19], and its schematic is depicted in Figure 1. The laser source was a homemade passively Q-switched Nd:YAG microchip laser. It can produce intense 1064-nm pulses with their duration of about 380 ps at 1 kHz. Various beam profiles can be achieved from this laser by changing the power, beam radius, and focus point of its pump. The linearly polarized laser beam was launched into the input face center of a GIMF by using an imaging system. The input patterns with different dominant LP modes of the GIMF were obtained by choosing similar laser beam profiles and adjusting the imaging system with three-axis translation stages. An isolator was used to protect the laser, and the input laser power could be adjusted by using a combination of a half wave plate and a polarizing beam splitter. A power meter was used to measure the reflective laser power from the PBS in order to estimate the input laser power. The GIMF (Nufern, GR-50/125-23HTA) has a core diameter of 50 μm and numerical aperture of 0.230. It should be noted that the bending condition of GIMF and thus linear coupling has great influences on the output beam profiles in our experiments. Therefore, the GIMF would be loosely coiled into circle of about 25 cm in diameter if there is no special statement. Output beam from the GIMF would enter different optical systems for investigating the evolutions of beam profiles and laser spectra, and details are shown in the following sections.

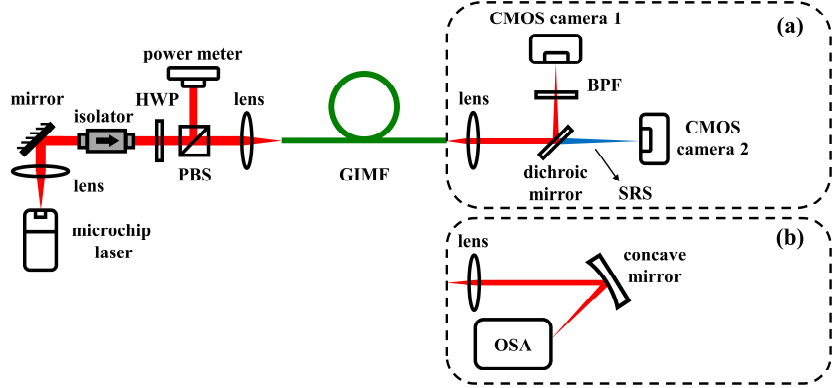

Figure 1. Experimental setup for the multimode laser propagation in a graded-index multimode fiber and the output beam characterization with the measurements for (a) near-field intensity spatial profiles at the laser and Raman wavelength, and (b) spectra. HWP: half wave plate; PBS: polarizing beam splitter; GIMF: graded-index multimode fiber; CMOS: complementary metal oxide semiconductor; BPF: band-pass filter; OSA: optical spectrum analyzer; SRS: stimulated Raman scattering.

In order to investigate the behavior of propagating multimode beam in the GIMF, we numerically solved the generalized multimode nonlinear Schrodinger equations using a massively parallel algorithm which is based on L. G. Wright's work [21]. Most of the laser and fiber parameters in the numerical simulations are the same as those in the experimental setup. The influences of nonlinear Kerr effects, stimulated Raman scattering, self-steepening and the first five order dispersion are included in the simulation. As a key factor in the NL-KBC and NL-BSM, disorder-induced linear coupling effect is also considered. In our simulations, the linear coupling disturbs the propagation by small, random, and symmetry-breaking index perturbations in the core of GIMF. These perturbations are used to simulate the inherent randomization which is introduced by the imperfections or bends in the GIMF. Moreover, limited by our computer performance, only the first twenty LP modes of the GIMF are considered. Further details of the numerical simulation are given in Section 2 of Supplementary Materials.

3. Results

In order to directly verify our conjecture about NL-BSM, the input beam profiles were adjusted to some specific cases similar to the high-order LP modes of GIMF. Here, the case with $LP_{12}$ mode as dominant one was first investigated, and its near-field beam profile is shown in Figure 2a. Due to its imperfect distribution compared to $LP_{12}$ mode, the input laser simultaneously excited other guided modes of GIMF. As shown in Figure 1a, after the propagation in a long GIMF, the near-field beam profile at the output face was imaged on a complementary metal-oxide-semiconductor camera (CINOGY, CinCam CMOS-1202) through a 25-mm aspherical lens with a magnification of about

15.5. A dichroic mirror highly refractive at 1064 nm and a 4-nm-wide band-pass interference optical filter at 1064 nm were used to analyze output beam profiles at the laser wavelength.

When the length of GIMF was 2 m, the evolution of output near-field beam profiles with the increasing input peak power is shown in Figure 2b-h, and more details are displayed in Supplementary Figure 1. At a relatively low peak power, such as 116.6 W in this case, the output beam intensity profile presented an irregular pattern. After the input laser power was increased to an appreciate level (see Figure 2d), the beam profile gradually evolved to a clear $LP_{12}$ mode structure surrounded by weak speckled background. Similar spatial beam evolution could be observed in those cases with the fiber length ranged from 1 m to 12 m, as shown in Supplementary Figure 2-6 and Visualization 1-4. These phenomena were similar to the NL-KBC effect [10, 19] where the $LP_{01}$ mode is the dominant mode of the input laser.

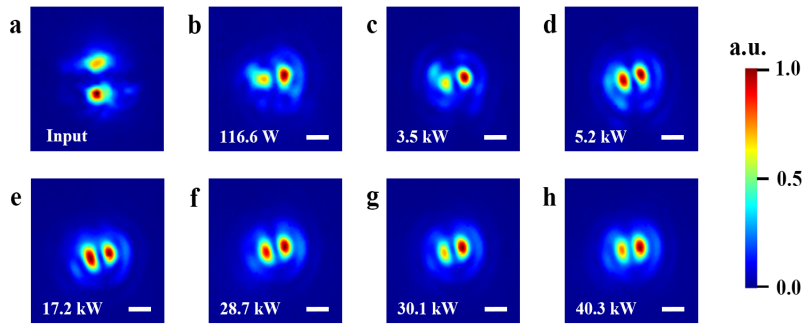

Figure 2. (a) The input beam profile and (b-h) the output near-field beam profile evolution from a 2-m GIMF with the increasing input peak power. The numerical value in the panel indicates the input peak power. Scale bars, 6.8 μm.

To investigate the physical mechanisms of this effect, we carried out numerical studies for multimode laser beam propagation in GIMF when $LP_{12}$ is the dominant mode. In this case, $LP_{12, b}$ mode accounts for about 91% of the total input energy, while the first nine LP modes only account for 1% each. The simulated results are presented in Figure 3. The simulated output beam intensity profiles in both linear and nonlinear regimes are in good agreement with our experimental results. Moreover, the numerical simulation clearly shows the evolution of laser transverse mode distribution along the GIMF, and hence the essence of NL-BSM. In linear regime when the input peak power is relatively low (about 210 W), the linear coupling introduced by the random index perturbations has great influences on the intermodal energy exchange. During the propagation, the proportion of dominant $LP_{12, b}$ mode decreases gradually from 91% to 40%, while the proportion of other modes slightly increases, especially $LP_{02}$ mode with the largest linear coupling coefficient to $LP_{12, b}$ mode. As a result, speckled output intensity profile can be achieved. In nonlinear regime when the input laser peak power is high enough (about 52.6 kW), the nonlinear effects overwhelm the influences of random linear coupling. Intermodal energy exchange from the dominant mode to other modes is greatly suppressed except for the degenerate $LP_{12, a}$ mode with the same propagation constant. As a result, although the proportion of $LP_{12, b}$ mode also decreases gradually to 37%, the proportion of degenerate $LP_{12, a}$ mode increases rapidly to 30%. Therefore, the proportion of dominant $LP_{12}$ mode (the combination of $LP_{12, a}$ and $LP_{12, b}$) are maintaining in a high level during the propagation, and hence the output beam intensity profile with a clear $LP_{12}$ mode structure can be achieved. This is why we call this phenomenon the nonlinear beam self-maintaining effect.

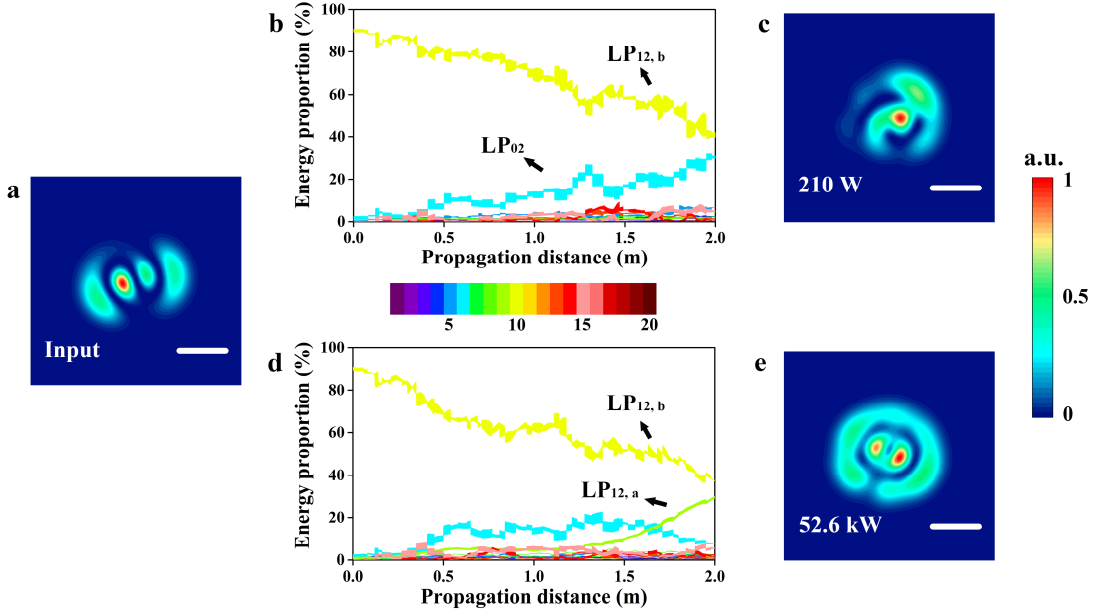

Figure 3. Numerical results of multimode beam propagation in a 2-m GIMF: (a) two-dimensional input beam intensity profile; the evolution of modal energy proportion along the propagation distance in the (b) linear and (d) nonlinear regimes; two-dimensional output beam intensity profiles in the (c) linear and (e) nonlinear regimes. The input peak power is about 210 W and 52.6 kW in linear and nonlinear regimes, respectively. The central color bar indicates the code of propagating LP modes. Scale bars, 10 μm.

Moreover, we imposed a mechanical deformation to the GIMF by manual squeezing, bending and twisting. As a result, a random and stronger linear coupling was introduced during the laser beam propagation. We recorded the evolution of output beam intensity profiles in the nonlinear regime when the input peak power was 5.2 kW (see Figure 2d and Visualization 5). Unlike the case of NL-KBC [10], the NL-BSM with input dominant $LP_{12}$ mode was sensitive to the deformation of the GIMF even in the nonlinear regime. These distinct performances can be attributed to the difference of propagation constants between the fundamental mode and HOMs [20]. It is also verified in our simulation that the NL-BSM is more robust to the disorder-induced mode coupling when the fundamental mode serves as the input dominant mode (see Supplementary Figure 14).

As shown in Figure 1b, the output spectral dynamics was investigated by using a commercial optical spectrum analyzer (Anritsu, MS9740A). Since a single-mode fiber was used in the optical input of this analyzer, only a small part of the HOMs can be collected (about 5% energy in this case). Therefore, the measured output spectrum can just qualitatively illustrate the relative intensity variation of each wavelength component. Figure 4a, b show that the output spectra followed a similar evolution to the case of NL-KBC reported in our previous work [19]. When the input peak power was increased, the output spectrum was almost unchanged at first, and then it was slightly broadened due to the self-phase modulation. With further increasing the laser power, the stimulated Raman scattering (SRS) and even super-continuum spectrum occurred,

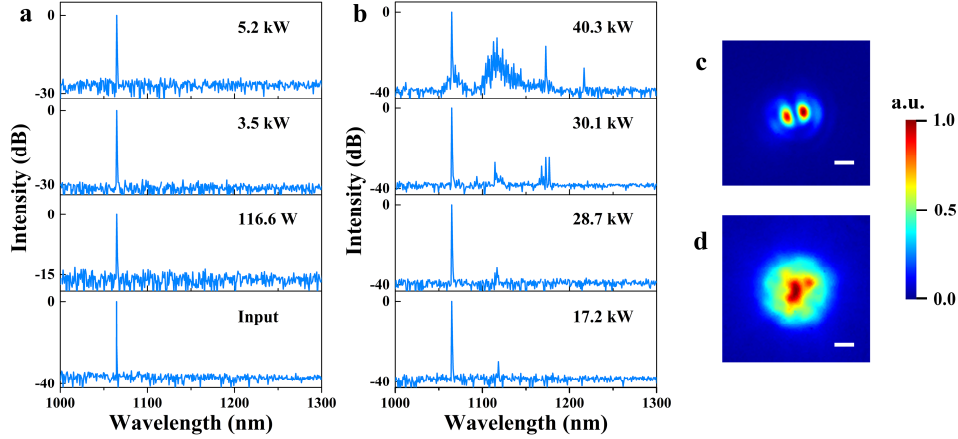

Figure 4. (a, b) The input and output spectra from a 2-m GIMF with the increasing launched peak power; the output near-field beam intensity profiles at (c) the laser wavelength and (d) the Raman wavelength when the input peak power was 28.7 kW. The numerical value in the panel indicates the input peak power. Scale bars, 6.8 μm.

According to Figure 2 and Figure 4, the obvious NL-BSM effect could be found (see Figure 2d) before the input peak power reached the SRS threshold (see Figure 4b, 17.2 kW in this case) or the catastrophic self-focusing threshold (in the megawatt region) [22]. It indicates that the driving mechanism of NL-BSM is probably the nonlinear Kerr effects but not SRS or the self-focusing. This is also verified by our numerical results in absence of Raman effect (see Supplementary Figure 15). With further increasing the laser power after the SRS occurred, the output beam profile kept a well-defined $LP_{12}$ mode structure at first, but then gradually evolved to an asymmetric pattern. In this process, the NL-BSM was weakened since the SRS caused serious dissipation of laser wavelength. As a result, the random intermodal energy exchange introduced by linear coupling regained the upper hand, and thus the distortion of output beam profile could be observed. Moreover, the output near-field beam profiles at the first-order SRS wavelength (1116.5 nm) were also recorded by the CMOS camera (Figure 1a). As shown in Figure 4d, a centered beam profile surrounded by strong multimode background could be observed, which verified the different performances between SRS beam cleanup [23] and NL-BSM.

For further investigating the mechanism of NL-BSM, we studied the spatial beam dynamics along the GIMF by backward cutting the fiber. Output beam profiles with similar input peak power of about 5.5 kW from the GIMF with different length are shown in Figure 5, and their complete spatial dynamics are presented in Supplementary Figure 1-6. In the nonlinear regime when the propagation length was short, multiple initial modes excited by the input beam and those modes with the large linear coupling coefficient could remain a relatively high energy level. Therefore, the comprehensive modal intensity distribution presented an asymmetric pattern (Figure 5a-c). After the laser pass through a longer distance, the beam profile gradually became a clearer $LP_{12}$ mode structure surrounded by weak speckled background (Figure 5d-f). We think these well-defined output beam profiles can be a synthetic effect caused by not only the nonlinear maintaining of input dominant mode but also the disorder-induced coupling of other modes. This phenomenon can also be found in our numerical results (see Supplementary Figure 13). However, since the number of propagating modes are limited in the simulation, the weak speckled background caused by disorder-induced coupling is not clear enough.

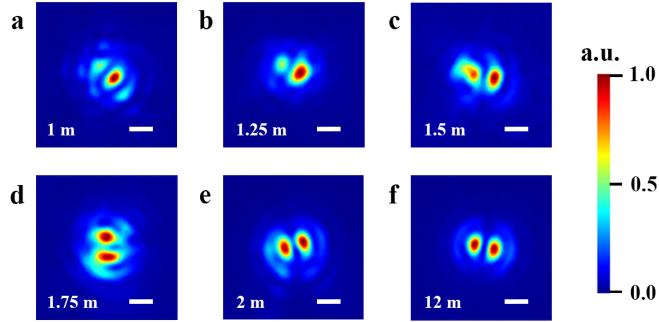

Figure 5. Experimental cut-back analysis of multimode beam propagation in a GIMF when the input peak power was about 5.5 kW. a-f show the output near-field beam intensity profiles (normalized intensity on a linear scale) with different propagation distances ranged from 1 m to 12 m. The rotation of the patterns can be due to fiber twist. Scale bars, 6.8 μm.

To find out the key factors for the NL-BSM effect, the laser beam (Figure 2a) was launched into a 1-m GIMF by imposing a transverse shift to the lateral position of the beam center with respect to the fiber core, while the output power was fixed. We recorded the evolution of output beam intensity profiles in the nonlinear regime when the input peak power was 44.24 kW (see Visualization 6). As shown in the video, with larger transverse shift and hence lower proportion of dominant $LP_{12}$ mode in the input beam, the NL-BSM effect was weaken and then completely absent. Similar phenomena can be found in the case of NL-KBC [6, 12] and our numerical results (see Supplementary Figure 16). These phenomena indicate that NL-BSM cannot be achieved if the input energy proportion of the dominant mode is too low.

We also investigated the NL-BSM effect of other HOMs in a 12-m GIMF. In addition to the $LP_{12}$ mode mentioned above, we targeted $LP_{11}$, $LP_{21}$, and $LP_{22}$ as the input dominant modes, respectively. The input and output near-field beam profiles are shown in Figure 6, and their complete spatial dynamics are presented in Supplementary Figure 6-8. In these cases, the evolution was similar: the output beam profile presented an irregular pattern with relatively low input laser peak power of about 100 W, but gradually evolved to a well-defined structure similar to the input dominant mode after the power was increased to a high level about 5 kW. These experimental results further verified the existence and universality of NL-BSM.

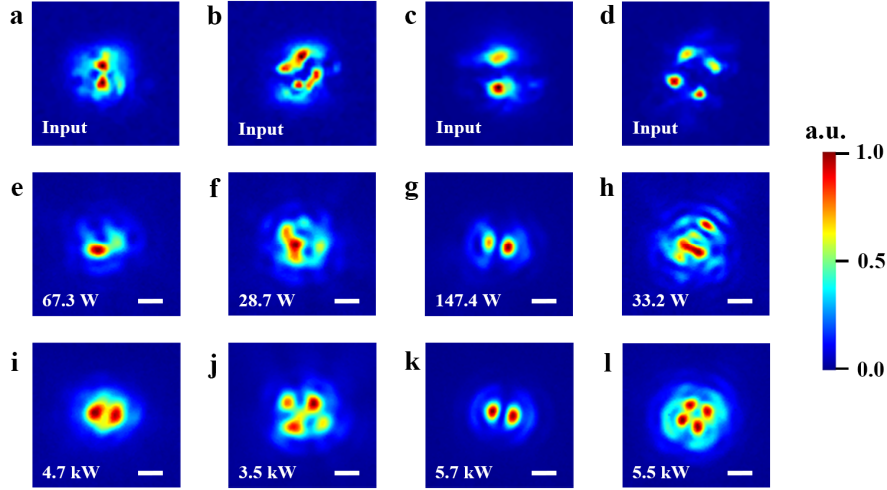

Figure 6. Experimental observation of nonlinear beam self-maintaining for different HOMs ($LP_{11}$, $LP_{21}$, $LP_{12}$, and $LP_{22}$ from left to right, respectively) when multimode beam propagation in a 12-m GIMF. a-l show the near-field intensity spatial profile of (a-d) input beam and output beam in the (e-h) linear and (i-l) nonlinear regimes, respectively. The numerical value in the panel indicates the input peak power. Scale bars, 6.8 μm.

## 4. Discussion

According to the experimental results above, we think NL-BSM is a typical self-organized process on which the nonlinearities and disorder have great influences. The performances and the mechanisms of the NL-BSM can be concluded below. When multimode laser beam with input dominant mode propagates in a long GIMF, mode coupling occurs between those modes with similar propagation constants. In practical situation, the disorder is introduced by the small random environmental perturbations and manufacturing errors to the fiber. This asymmetric disorder can lead to similar propagation constants and hence large mode-coupling coefficients of neighboring modes [20]. Therefore, intermodal energy exchange occurs, and a beam profile with irregular pattern can be achieved after a propagation distance in linear regime. In nonlinear regime, the index perturbation caused by the Kerr nonlinearity is vastly different between the dominant mode and other modes due to their energy gap. As a result, the propagation constant of input dominant mode is different from other modes except its degenerate mode(s). Therefore, the energy of input dominant mode can be maintaining during the propagation while other modes still suffer from intermodal energy exchange due to the disorder. It should be noted that the energy of input dominant mode can be maintaining (slightly decreased at first, see Supplementary Figure 13) but cannot be enhanced by comparison to the input condition during the propagation. Moreover, the key factors for the existence of NL-BSM can be also summarized: the propagation conditions produced by GIMF, the appropriate linear coupling due to small disorder, the propagation constant differences introduced by nonlinear Kerr effects, and the input beam with dominant mode.

The threshold of NL-BSM is another point of interest when applying this effect. Since the Kerr nonlinearity is the driving mechanism of NL-BSM, there should be no strict threshold. However, distinguished by the phenomena, we can define two power thresholds of NL-BSM for a specific input and propagation condition. The lower threshold is defined as the power at which the nonlinear Kerr effects overwhelm the influences of random linear coupling, and hence a clear output beam profile similar to input dominant mode can be observed. The upper threshold is defined as the power at which the output beam profile becomes asymmetric pattern again due to the SRS dissipation. These two thresholds can be achieved by calculating the correlation of the output intensity profiles with the input dominant mode.

Comparing our experimental results with the NL-KBC effect reported in previous works [10, 19], we can find out plenty of similarities between the NL-KBC and NL-BSM. Therefore, we here classify the NL-KBC as an exceptional case of NL-BSM in which the fundamental mode serves as the input dominant mode. Moreover, although the NL-BSM of $LP_{11}$, $LP_{21}$, $LP_{12}$, and $LP_{22}$ modes have been experimentally demonstrated, we do not think NL-BSM can be achieved with any input dominant mode. According to the structure of GIMF, higher-order modes have more neighbors with similar propagation constants, and hence they are more sensitive to the disorder-induced mode coupling. The disorder introduced by the fiber manufacturing errors will be strong enough to distort the NL-BSM for HOMs. Therefore, in a practical situation, the NL-BSM may be limited to some relatively low-order modes.

The limitation of our experiments and numerical simulations will also be discussed. In the experiments, we only obtained a few HOMs which are restricted by our laser design. In addition, these input beam profile were in a low quality by comparison to the propagation modes of GIMF. As a result, the proportion of input dominant mode was low, for example $LP_{21}$ mode in our case, and hence the output beam profiles introduced by the NL-BSM are slightly distorted. In the numerical simulations, although most of the laser and fiber parameters are the same as those in our experiments, the input beam profiles are artificially designed. Moreover, limited by our computer performance, only the first twenty LP modes of the GIMF are considered, and hence mode coupling of higher-order modes is distorted in the simulations. In addition, linear losses resulted by the coupling to cladding modes are not considered. All these limitations lead to the main differences between the experimental and numerical results.

## 5. Conclusion

We demonstrated experimentally the existence of NL-BSM when intense laser beam with dominant mode propagates through a long GIMF. During the propagation, the random linear coupling introduced by the disorder and the mode coupling introduced by nonlinear Kerr effects have great influences on the intermodal energy exchange. As a result, the output beam profile presents an irregular pattern in linear regime, while it becomes a well-defined structure since the energy of input dominant mode is maintaining in nonlinear regime. In addition to the performances, the key factors and the universality of NL-BSM have also been investigated. According to the experimental and numerical results, the NL-KBC is classified as an exceptional case of NL-BSM. We hope that our works can also provide a new perspective to understand other multimode fiber nonlinearities, especially multimode solitons and spatiotemporal mode-locking. Future works will focus on the nonlinear mechanisms and the applications of this NL-BSM effect.

**Acknowledgements**
The authors would like to acknowledge Zhanwei Liu for the helpful discussion on the numerical model and the MPA program. We acknowledge Fujuan Wang and Jiaoyang Li for the technical support in this work. This work was partially supported by National Natural Science Foundation of China (Grant No. 11832019) and the NSFC Original Exploration Project (Grant No. 12150001).

**Disclosures**

The authors declare no conflicts of interest.

# Supplementary materials for "Nonlinear beam self-maintaining effect in graded-index multimode fiber"

Baofu Zhang[1], Shanchao Ma[2], Qiurun He[2], Sihua Lu[2], Wei Guo[2], Zhongxing Jiao[2, *], and Biao Wang[1, 2, *]

[1] Research Institute of Interdisciplinary Science & School of Materials Science and Engineering, Dongguan University of Technology, Dongguan 523808, China

[2] School of Physics, Sun Yat-sen University, Guangzhou 510275, China

*Corresponding author: jiaozhx@mail.sysu.edu.cn, wangbiao@mail.sysu.edu.cn.

## 1. Supplementary experimental results

In order to clearly verify the nonlinear beam self-maintaining effect (NL-BSM) with different input conditions, supplementary experimental results are shown in Section 1.

Here, the case with dominant $LP_{12}$ mode was first investigated. Supplementary Figure 1-6 show the output near-field beam profile evolution with the increasing input peak power in a graded-index multimode fiber (GIMF) by backward cutting the fiber with its length ranged from 12 m to 1 m. The rotation of the output patterns was different between these cases with different fiber length, which was caused by the fiber twist. Moreover, these beam profile evolutions are also shown in Visualization 1 (corresponding to Supplementary Figure 3), Visualization 2 (corresponding to Supplementary Figure 4), Visualization 3 (corresponding to Supplementary Figure 5), and Visualization 4(corresponding to Supplementary Figure 6).

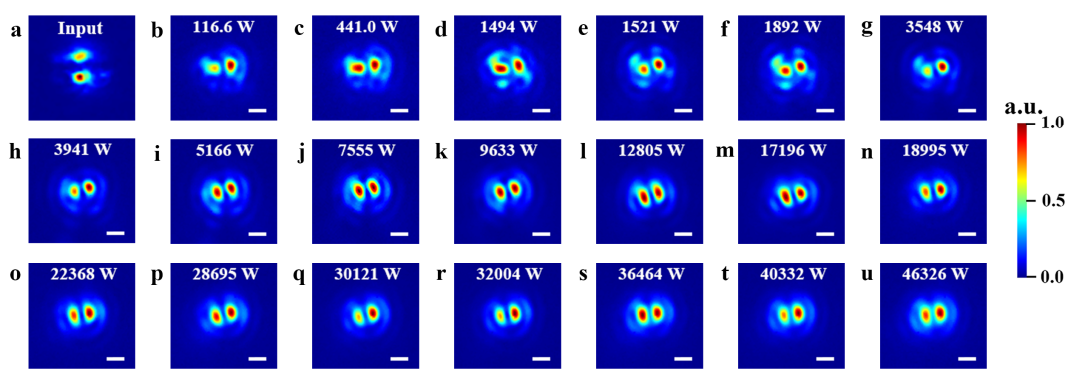

Supplementary Figure 1: (a) The input beam profile and (b-u) the output near-field beam profile evolution from a 2-m GIMF with the increasing input peak power. The numerical value in the panel indicates the input peak power. Scale bars, 6.8 μm.

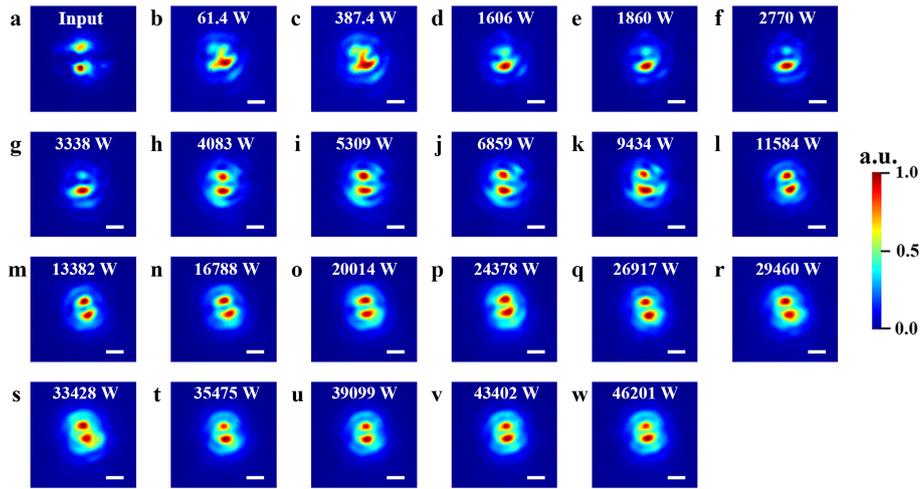

Supplementary Figure 2: (a) The input beam profile and (b-w) the output near-field beam profile evolution from a 1.75-m GIMF with the increasing input peak power. The numerical value in the panel indicates the input peak power. Scale bars, 6.8 μm.

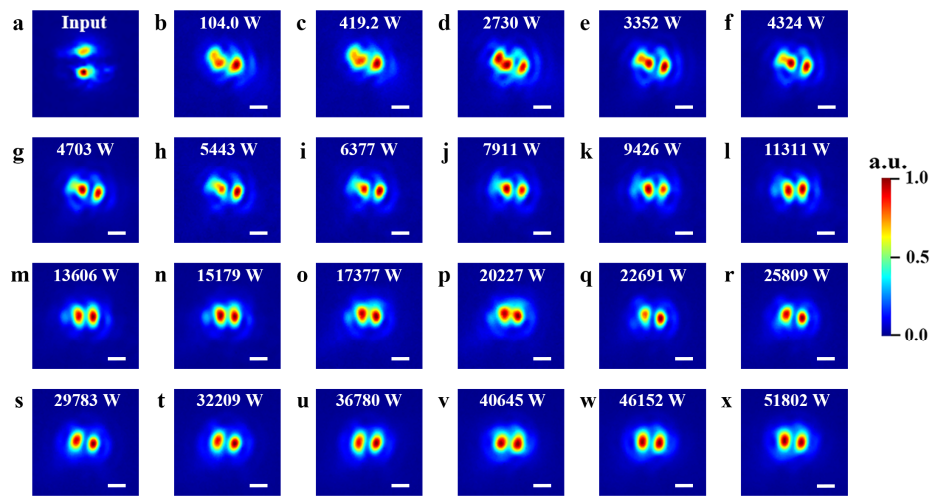

Supplementary Figure 3: (a) The input beam profile and (b-x) the output near-field beam profile evolution from a 1.5-m GIMF with the increasing input peak power. The numerical value in the panel indicates the input peak power. Scale bars, 6.8 μm.

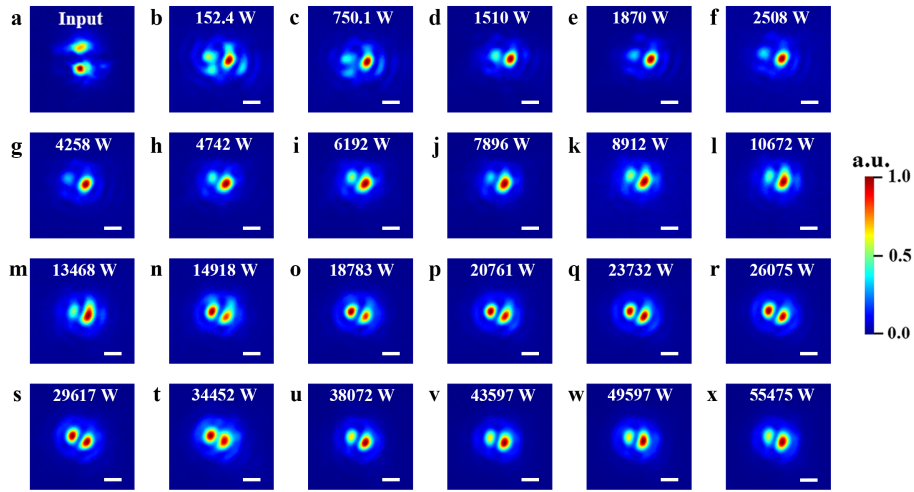

Supplementary Figure 4: (a) The input beam profile and (b-x) the output near-field beam profile evolution from a 1.25-m GIMF with the increasing input peak power. The numerical value in the panel indicates the input peak power. Scale bars, 6.8 μm.

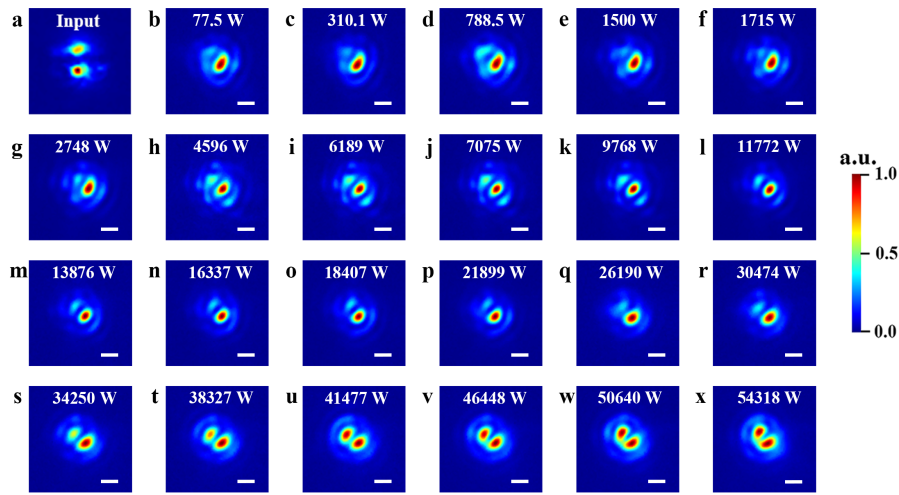

Supplementary Figure 5: (a) The input beam profile and (b-x) the output near-field beam profile evolution from a 1-m GIMF with the increasing input peak power. The numerical value in the panel indicates the input peak power. Scale bars, 6.8 μm.

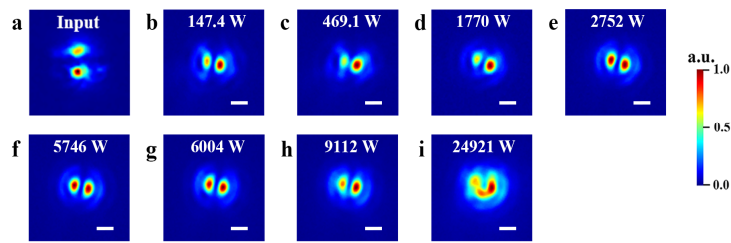

Supplementary Figure 6: (a) The input beam profile similar to $LP_{12}$ mode and (b-i) the output near-field beam profile evolution from a 12-m GIMF with the increasing input peak power. The numerical value in the panel indicates the input peak power. Scale bars, 6.8 μm.

Supplementary Figure 6-8 show nonlinear beam self-maintaining effect for different dominant HOMs ($LP_{12}$, $LP_{11}$, and $LP_{21}$) in a 12-m GIMF. The output beam profiles for the cases of $LP_{11}$ and $LP_{21}$ show lower quality by comparison to the case of $LP_{12}$. This may be caused by the lower proportion of dominant mode in those two cases. Moreover, the laser crystal was damaged by the intense pump beam during the experiment with the input beam profile similar to $LP_{22}$ mode. Therefore, only the input beam profile, typical output beam profiles in linear and nonlinear regime were recorded and shown in the article.

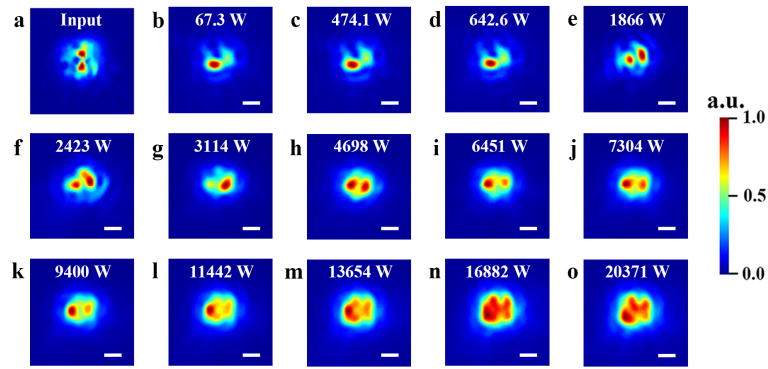

Supplementary Figure 7: (a) The input beam profile similar to $LP_{11}$ mode and (b-o) the output near-field beam profile evolution from a 12-m GIMF with the increasing input peak power. The numerical value in the panel indicates the input peak power. Scale bars, 6.8 μm.

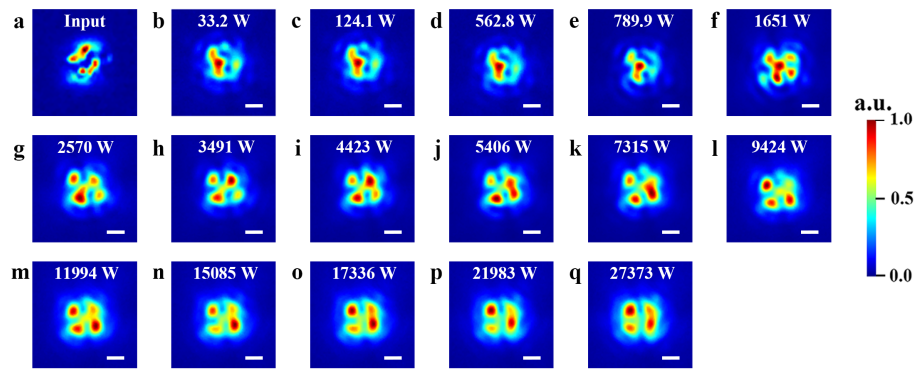

Supplementary Figure 8: (a) The input beam profile similar to $LP_{21}$ mode and (b-q) the output near-field beam profile evolution from a 12-m GIMF with the increasing input peak power. The numerical value in the panel indicates the input peak power. Scale bars, 6.8 μm.

Supplementary Figure 9 shows the output spectral evolution with the increasing input peak power corresponding to Supplementary Figure 1-5. Supplementary Figure 10 shows the output spectral evolution from a 12-m GIMF with the increasing input peak power corresponding to Supplementary Figure 6-8.

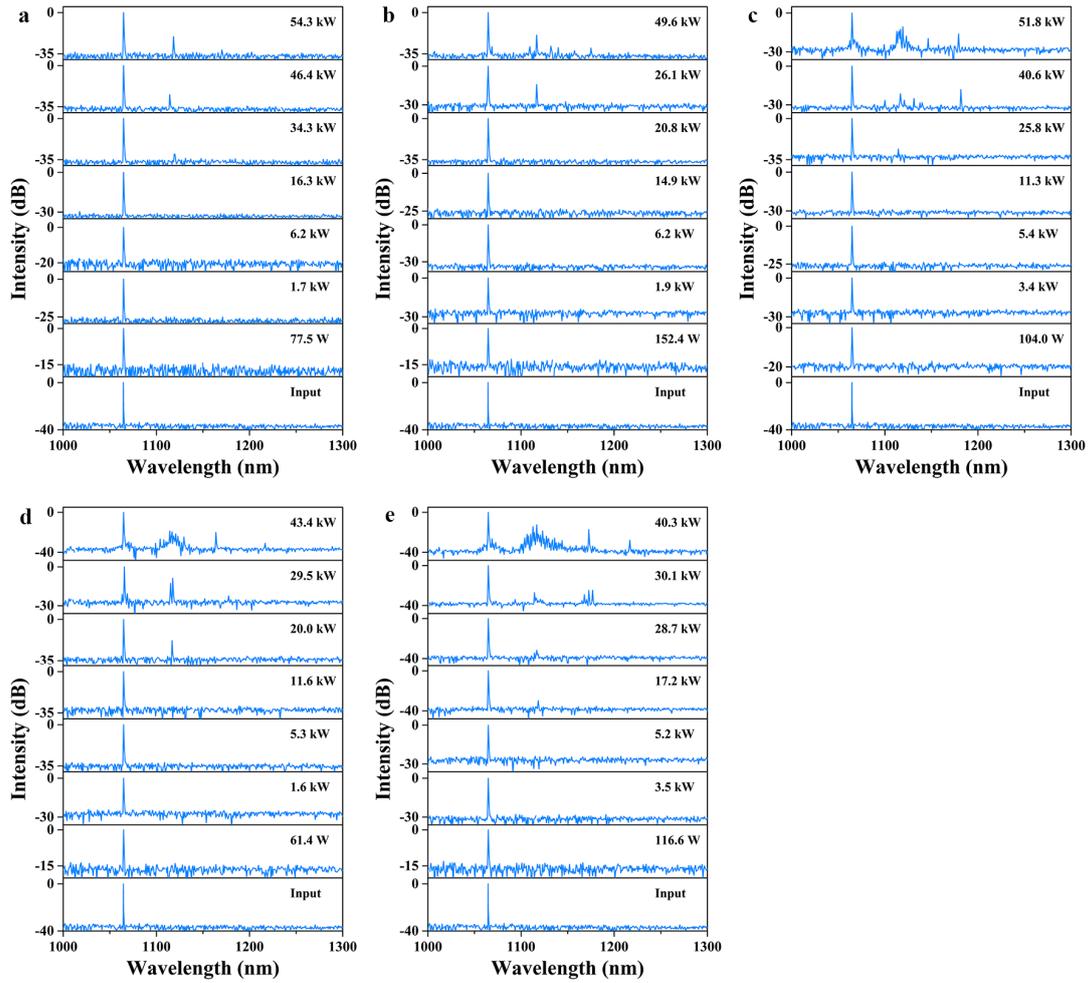

Supplementary Figure 9: The input and output spectra from the (a) 1-m, (b) 1.25-m, (c) 1.5-m, (d) 1.75-m, and (e) 2-m GIMF with the increasing input peak power. The numerical value in the panel indicates the input peak power.

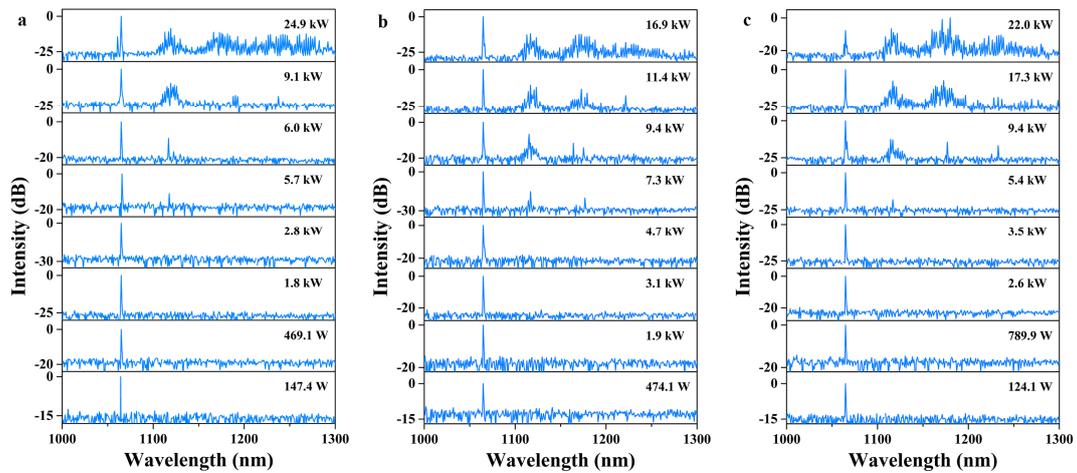

Supplementary Figure 10: The output spectra of the multimode laser beam with dominant (a) $LP_{12}$ mode, (b) $LP_{11}$ mode, and (c) $LP_{21}$ mode from a 12-m GIMF with the increasing input peak power. The numerical value in the panel indicates the input peak power.

## 2. Numerical method

In order to simulate multimode beam propagation in GIMF and investigate the NL-BSM, we numerically solved the generalized multimode nonlinear Schrodinger equations (GMMNLSE). The GMMNLSE we solved here are a set of coupled NLSE-type equations to describe the dynamics of the electric field temporal envelope $A_p(z, t)$ for spatial mode $p$ [1, 2]:

$$\partial_z A_p(z,t) = i(\beta_0^{(p)} - \text{Re}[\beta_0^{(0)}])A_p - (\beta_1^{(p)} - \text{Re}[\beta_1^{(0)}])\partial_t A_p + \sum_{m=2}^{M} i^{m+1} \frac{\beta_m^{(p)}}{m!} \partial_t^m A_p + i\sum_{q \neq p}^{N} C_{pq}(z) A_q$$
$$+ i\frac{n_2 \omega_0}{c}\left(1 + \frac{i}{\omega_0}\partial_t\right) \sum_{l,m,n}^{N} \left\{(1-f_R) S_{plmn}^K A_l A_m A_n^* + f_R S_{plmn}^R A_l \int_{-\infty}^{t} \left[h_R(\tau) A_m(z,t-\tau) A_n^*(z,t-\tau)\right]d\tau\right\},$$
(S1)

where $\omega_0$ is the carrier angular frequency.

The first three terms on the right-hand side of equation (S1) yield the effects of dispersion for mode $p$ by a Taylor series expansion $\beta_m^{(p)} = \partial^m \beta^{(p)}/\partial\omega^m$ about $\omega_0$, then transforming the terms into the time domain. Here the modal propagation constants $\beta^{(p)}$ are complex values where the imaginary part describes mode and wavelength dependent losses; Re[..] denotes the real part only. Therefore, the first two terms are expressed relative to the lowest-order longitudinal phase evolution and group velocity of the fundamental mode. The third term represents higher-order dispersion effects, and only the first five order dispersion is considered in our simulations.

The fourth term on the right hand side of equation (S1) represents linear mode coupling effect including disorder [2-4]. This term will be ignored in most cases but it is very important for NL-BSM. The coupling coefficients $C_{pq}(z)$ can be given by [2-4]:

$$C_{pq}(z) = \frac{k_0}{2n_{eff}} \cdot \frac{\int dxdy [n^2(x,y,z) - n_p^2(x,y,z)] F_q(x,y) F_p^*(x,y)}{[\int dxdy F_q^2 \int dxdy F_p^2]^{1/2}},$$
(S2)

where $n_p(x, y, z)$ is the perturbed index profile, $n(x, y, z)$ is the ideal profile, and $F_p(x, y)$ is a discrete transverse fiber mode profile. In our simulations, we assume that the disorder is introduced by random bends of the fiber, and one of them can be described by [5]:

$$\Delta n(x,y,z_n) = n_p(x,y,z_n) - n(x,y,z_n) = \left(\frac{\Delta_x x}{R} + \frac{\Delta_y y}{R}\right)_{z_n}$$
(S3)

for $\sqrt{x^2 + y^2} \leq R$ and $\Delta n(x, y, z_n) = 0$ otherwise. The values $\Delta_x$ and $\Delta_y$ are random numbers obtained from a uniform distribution ranging from $-a * \Delta$ to $a * \Delta$, where $\Delta$ is the difference between the center and cladding index of the fiber, coefficient $a$ indicates the disorder level. In our simulations, $a$ is set to be 1.6% if there is no specified statement. In order to ensure the aperiodic disorder, the values $\Delta_x$ and $\Delta_y$ are changed every $z_n$ cm, where $z_n$ is a random number from a uniform distribution [2,6] except the last propagation distance [5]. It should be noted that the distance $z_n$ is very long compared to intergroup beat lengths in the GIMF and the basic integration step in our simulations.

The second line of equation (S1) represents the nonlinear optical effects with a nonlinear refractive index $n_2$. The term $\propto \partial/\partial t$ describes self-steepening and the two terms within the sum describe Kerr and Raman nonlinearities [1]. $h_R$ is the Raman response of the fiber medium, $f_R$ is the Raman contribution to the Kerr effect ($f_R = 0.18$ for silica glass fiber and in our case). $S_{plmn}^R$ and $S_{plmn}^K$ are the nonlinear coupling coefficients for the Raman and Kerr effect respectively [1]. In our simulations, by assuming that the modes excited are in a single linear polarization, and neglecting

spontaneous processes, $S^R_{plmn}$ and $S^K_{plmn}$ can be given by [1]:

$$S^R_{plmn} = S^K_{plmn} = \frac{\int dxdy[F_p F_l F_m F_n]}{[\int dxdy F_p^2 \int dxdy F_l^2 \int dxdy F_m^2 \int dxdy F_n^2]^{1/2}}, \tag{S4}$$

In order to solve the GMMNLSE, we used a numerical method called massively parallel algorithm (MPA) which is based on L. G. Wright's work [2], and their numerical solver is freely available online [6]. Disorder has been considered in our simulations, and hence the fourth term on the right hand side of equation (S1) is the main difference between the original numerical solver online [6] and our modified version. Since the detail of MPA has been systematically demonstrated in the Reference [2] and the documentation of solver online [6], we will not repeat the algorithm but present our parameters below. In our simulations, we used an integration step of 25 μm, a transverse 800 × 800 grid for a spatial window of 125 μm × 125 μm, and the time grid points of $2^{15}$ for a time window of 1500 ps. A typical GIMF with a truncated parabolic-like profile of the refractive index in the transverse domain is considered in the simulations. The GIMF has a core radius of $R$ = 25 μm, a maximum value of $n_{co}$ = 1.4633 for the core refractive index, a cladding radius of 62.5 μm, and a cladding index of $n_{cl}$ = 1.4496. Thus, $n(x,y,z) = n_{co} - \Delta(\sqrt{x^2+y^2}/R)^\alpha$ for $\sqrt{x^2+y^2} \leq R$ and $n(x, y, z) = n_{cl}$ otherwise, where α = 2.08 and Δ = $n_{co}$ - $n_{cl}$ = 0.0137. In our simulations, we use $n_2$ = 2.3 × $10^{-20}$ m$^2$W$^{-1}$ and $f_R$ = 0.18. In some situations, stimulated Raman scattering can be ignored by letting $f_R$ = 0. If there is no specified statement, we use typical fiber and laser parameters which are similar to our experimental setup: the propagation length of 2 m, input laser beam with the total pulse energy of 20000 nJ and a pulse duration of 380 ps (that is, a peak power of 52.6 kW). Limited by our computer performance, only the first twenty LP modes of the GRIN MMF are considered in the simulations. These modes are, in order, $LP_{01}$, $LP_{11,a}$, $LP_{11,b}$, $LP_{21,a}$, $LP_{21,b}$, $LP_{02}$, $LP_{31,a}$, $LP_{31,b}$, $LP_{12,a}$, $LP_{12,b}$, $LP_{41,a}$, $LP_{41,b}$, $LP_{22,a}$, $LP_{22,b}$, $LP_{03}$, $LP_{41,a}$, $LP_{41,b}$, $LP_{32,a}$, $LP_{32,b}$, and $LP_{13,a}$. Their beam intensity profiles are shown in Supplementary Figure 11. If there is no specified statement, we use a typical input beam profile in which $LP_{12,b}$ mode accounts for about 91 % of the total input energy, while the first nine LP modes account for 1% each.

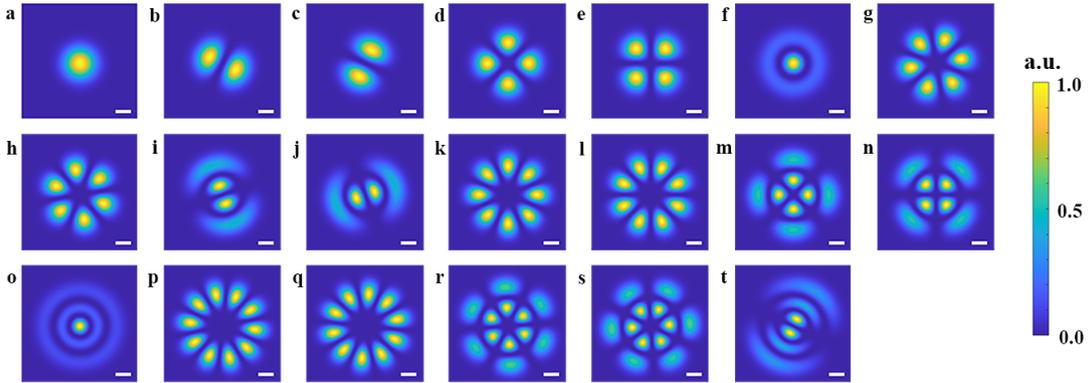

Supplementary Figure 11: beam intensity profiles of first twenty propagating LP modes. a-t are, in order, $LP_{01}$, $LP_{11,a}$, $LP_{11,b}$, $LP_{21,a}$, $LP_{21,b}$, $LP_{02}$, $LP_{31,a}$, $LP_{31,b}$, $LP_{12,a}$, $LP_{12,b}$, $LP_{41,a}$, $LP_{41,b}$, $LP_{22,a}$, $LP_{22,b}$, $LP_{03}$, $LP_{41,a}$, $LP_{41,b}$, $LP_{32,a}$, $LP_{32,b}$, and $LP_{13,a}$. Scale bars, 5 μm.

## 3. Supplementary simulated results

For further investigating the mechanism of NL-BSM, supplementary simulated results are shown in Section 3. Supplementary Figure 12 shows color scale for the code of propagating LP modes in the simulated results of this section.

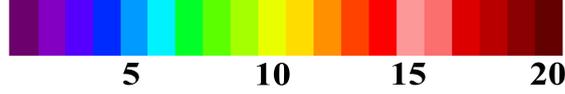

Supplementary Figure 12: Color scale for the code of propagating LP modes.

Beam profiles in any position of GIMF and the evolution of modal energy proportion along the propagation distance can be achieved in the simulations. Supplementary Figure 13 shows typical numerical results of laser beam with dominant $LP_{12, b}$ mode propagation in a 2-m GIMF in the nonlinear regime. As shown in this figure, the energy proportion of dominant $LP_{12}$ mode ($LP_{12, a}$ + $LP_{12, b}$) decreases at first, and then it can be maintaining in a high level of about 65%.

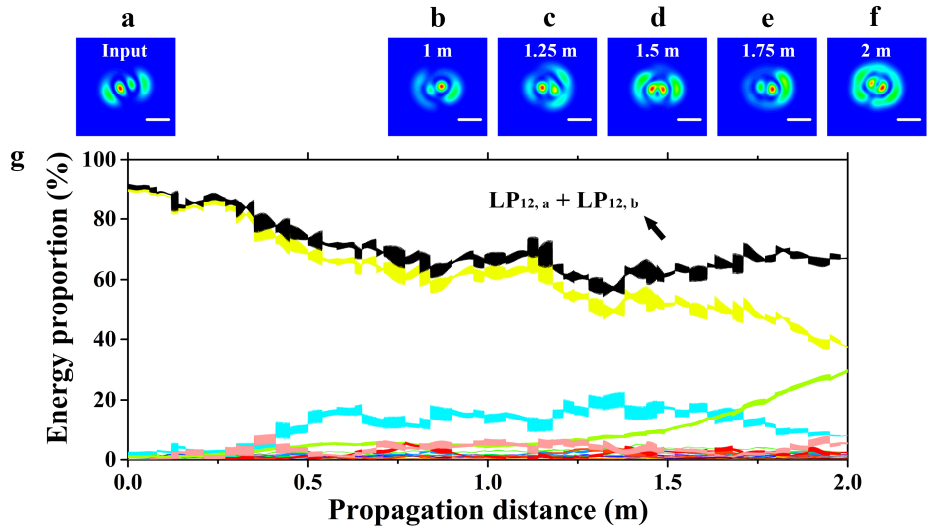

Supplementary Figure 13: Numerical results of laser beam with dominant $LP_{12, b}$ mode propagation in a 2-m GIMF in the nonlinear regime: (a-f) two-dimensional propagating beam intensity profiles in different positions, and (g) the evolution of modal energy proportion along the propagation distance. The input peak power is about 52.6 kW. Scale bars, 10 μm.

When a mechanical deformation is imposed to the GIMF, a stronger linear coupling was introduced during the laser beam propagation. Supplementary Figure 14 shows the numerical results of laser beam with dominant $LP_{01}$ mode and LP $LP_{12, b}$ mode propagation in a 2-m GIMF with a stronger disorder in the nonlinear regime. In these simulations, the coefficient *a* in Section 2 is set to be a larger value of 2.5% in order to evaluate a stronger disorder. The simulated results indicate that the NL-BSM for the $LP_{12}$ mode is sensitive to the disorder-induced mode coupling while the NL-BSM for $LP_{01}$ mode is more robust. This is in a good agreement with our experimental results (Visualization 5) and the case of nonlinear Kerr beam self-cleaning with similar input peak power [7].

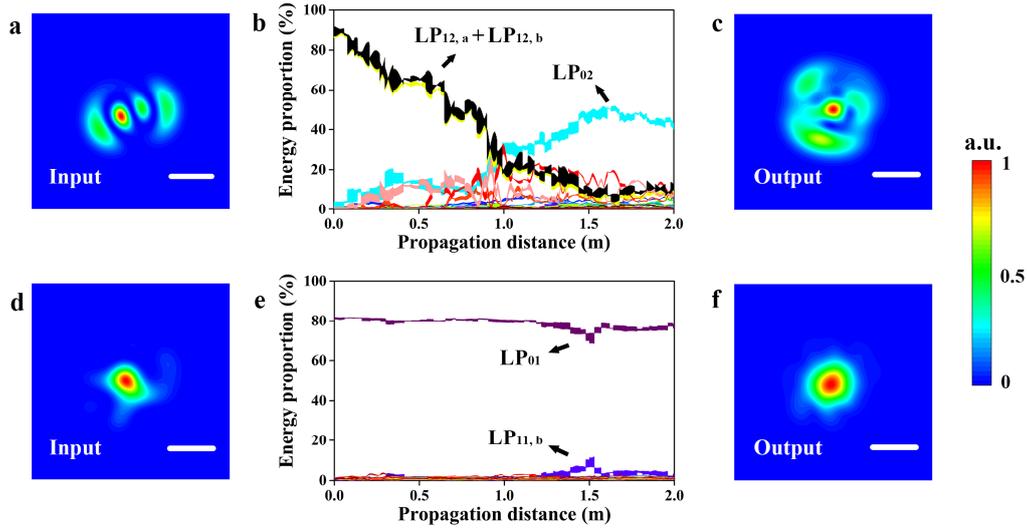

Supplementary Figure 14: Numerical results of laser beam with dominant (a-c) $LP_{12,b}$ mode and (d-f) $LP_{01}$ mode propagation in a 2-m GIMF in the nonlinear regime: two-dimensional (a, d) input and (c, f) output beam intensity profiles, and (b, e) the evolution of modal energy proportion along the propagation distance. The input peak power is about 52.6 kW. Scale bars, 10 μm.

In order to figure out the physical mechanism of NL-BSM, the simulation in absence of Raman effect was carried out, and the numerical results are shown in Supplementary Figure 15. These results indicate that the NL-BSM can be achieved even if the Raman effect is neglecting. This provides an indirect evidence for the conjecture that the driving mechanism of NL-BSM is probably the nonlinear Kerr effects.

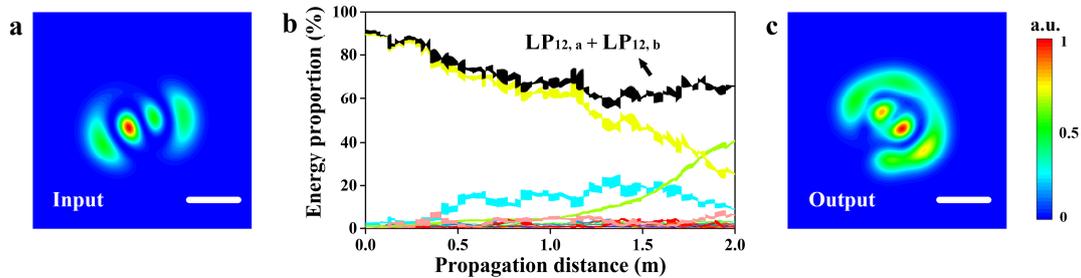

Supplementary Figure 15: Numerical results of laser beam with dominant $LP_{12,b}$ mode propagation in a 2-m GIMF in the nonlinear regime but in absence of Raman effect: two-dimensional (a) input and (c) output beam intensity profiles, and (b) the evolution of modal energy proportion along the propagation distance. The input peak power is about 31.6 kW. Scale bars, 10 μm.

When a transverse shift is imposed to the lateral position of the input beam center with respect to the fiber core, the energy proportion of dominant $LP_{12,b}$ mode will decrease, and more high-order modes can be excited. In the simulations, the energy proportion of dominant $LP_{12,b}$ mode in the input beam is set to be a smaller value to evaluate a larger transverse shift, while the first nine LP modes split the rest of input energy equally. Supplementary Figure 16 shows the numerical results of laser beam propagation in a 2-m GIMF in the nonlinear regime, with the input energy proportion of $LP_{12,b}$ mode ranged from 64% to 91%. These simulated results indicate that the energy of the input dominant mode can be maintaining but not enhanced during the propagation. Moreover, NL-BSM cannot be achieved if the input energy proportion of the dominant mode is too low. This is in a good agreement with our experimental results (Visualization 6) and the case of nonlinear Kerr beam self-cleaning [8].

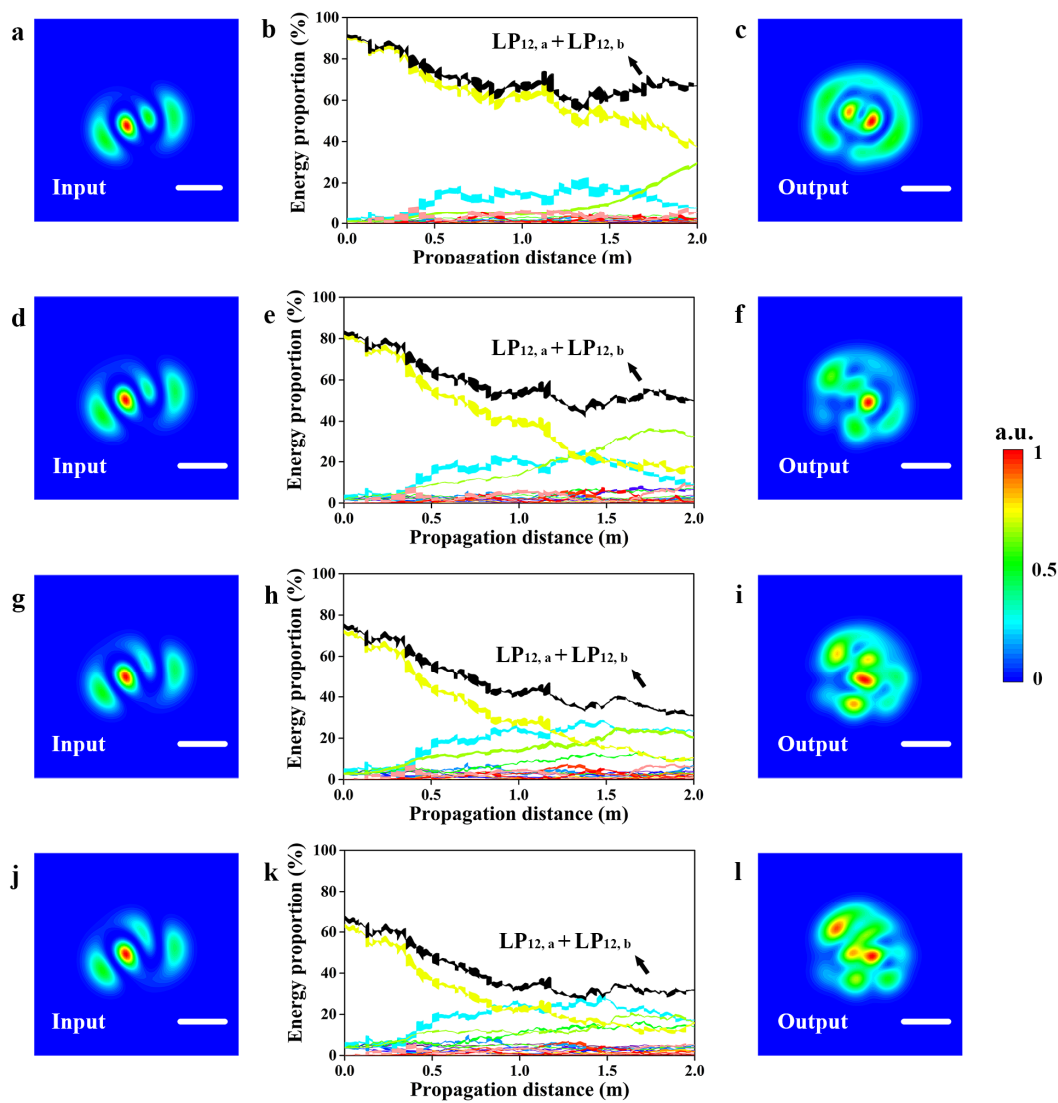

Supplementary Figure 16: Numerical results of laser beam with different input condition propagation in a 2-m GIMF in the nonlinear regime: two-dimensional (a, d, g, j) input and (c, f, I, l) output beam intensity profiles, and (b, e, h, k) the evolution of modal energy proportion along the propagation distance. The $LP_{12,b}$ mode accounts for (a, b, c) 91%, (d, e, f) 82%, (g, h, i) 73%, and (j, k, lc) 64% of the total input energy, respectively. The input peak power is about 52.6 kW. Scale bars, 10 μm.

**Reference for Supplementary Material**